\documentstyle[aps,prl,epsfig,preprint]{revtex}
\begin{document}

\title{ 
Electronic non-adiabatic states}

\author{Nikitas I. Gidopoulos$^1$ and E.K.U. Gross$^2$ \\
$^1$ISIS Facility, Rutherford Appleton Laboratory, 
Chilton, Didcot, Oxon, OX11 0QX, England, U.K.\\
$^2$Institut f\"ur Theoretische Physik, Freie Universit\"at Berlin, 
Arnimallee 14, D-14195 Berlin, Germany 
}
\date{\today}
\maketitle
\begin{abstract}
A novel treatment of non-adiabatic couplings is proposed. The derivation 
starts from the long-known, but not well-known, fact that the wave function 
of the complete system of elctrons and nuclei can be written, without 
approximation, as a Born-Oppenheimer-type product of a nuclear wavefunction, 
$X(R)$, and an electronic one, $\Phi_R(r)$, which depends parametrically on 
the nuclear configuration $R$. From the variational principle we deduce 
formally exact equations for $\Phi_R(r)$ and $X(R)$. The algebraic structure 
of the exact nuclear equation coincides with the corresponding one in the 
adiabatic approximation. The electronic equation, however, contains terms not 
appearing in the adiabatic case, which couple the electronic and the nuclear 
wavefunctions and account for the electron-nuclear correlation beyond the 
Born-Oppenheimer level. It is proposed that these terms can be incorporated 
using an optimized local effective potential.
\end{abstract}

\section{Introduction}

The Born Oppenheimer (BO) \cite{bo,bh} or adiabatic \cite{longuet} separation 
of the electronic and nuclear motion forms the basis of almost any quantum 
mechanical study of molecules and solids.  The method is based on the large 
difference between the electronic and nuclear masses, which implies that the 
light electrons are able to adapt almost instantaneously to the configuration 
of the much heavier nuclei.  Consider a molecule with $N_{\rm e}$ electrons 
and $N$ nuclei with masses $M_{\alpha}$ and charges $Z_{\alpha}$ and denote by 
$r$ the set of electronic coordinates $r = {\bf r}_1 \ldots {\bf r}_{N_{\rm e}}$ 
and by $R$ the set of nuclear coordinates $R = {\bf R}_1, \ldots {\bf R}_N$.  

The separation is effected by decomposing the complete molecular wavefunction, 
$ \Psi_{\rm mol}(r,R)$, in terms of a product of an electronic wavefunction 
depending parametrically on the nuclear positions and a nuclear wavefunction:
\begin{equation}
\Psi_{\rm mol}(r,R) \simeq  \Phi_R^{\rm BO}(r) \, X^{\rm BO}(R)  .
\end{equation}

Since the electrons, more or less, cannot see the nuclear motion, the 
electronic wavefunction is chosen to be an eigenfunction of an electronic 
Hamiltonian derived from the molecular one by ignoring the nuclear kinetic 
energy or, equivalently, the nuclear mass is set to infinity. Finally, 
following Longuet-Higgins \cite{longuet} and Messiah \cite{messiah}, the 
equation for the nuclear wavefunction can be obtained variationally.  
Keeping the electronic wavefunction fixed, one requires that the expectation 
value of the molecular Hamiltonian in terms of the adiabatic product 
wavefuction remains stationary when the nuclear wavefunction is varied.  
In the resulting effective nuclear Hamiltonian, in the absence of electronic 
degeneracies, the nuclei lie in a potential which, to high accuracy, coincides 
with the electronic energy eigenvalue.  Each  electronic level allows for a 
set of nuclear vibrational and rotational excitations but does not couple 
directly to any nuclear state of this set. For many physical systems and 
phenomena this separation is extremely successful, at least whenever degeneracy 
or near degeneracy of the electronic Hamiltonian is not involved.  There are 
however other phenomena, which crucially depend on the coupling between the 
electronic and the nuclear wavefunctions. For these phenomena the adiabatic or 
BO approximation breaks down.  

One should be careful about the variational derivation of the nuclear BO 
equation, because the ground state of a free molecule is a continuum state 
(since the complete molecular Hamiltonian is translationally invariant). 
Rigorously then, one should first separate off the centre of mass of the 
molecule and then perform the adiabatic decomposition of the resulting bound 
wavefuction \cite{werner}.  This leads to complicated terms in the resulting 
Hamiltonian.  In this paper we do not separate off the centre of mass first, 
because our main aim is to show how one may improve upon the BO scheme, keeping 
at the same time the formulae simple and transparent, rather than maintaining 
complete rigour.  One may justify the approach by imposing a confining potential 
that bounds the molecule within a very large but finite volume.  

In probability theory, the joint probability distribution $p(r,R)$ of two sets 
of variables $r, R$ can be factorized as the product $p(r,R) = p(r|R) \, p(R)$ 
of a conditional probability distribution $p(r|R)$ for the (electronic) variables 
$r$ given $R$, times a marginal probability distribution $p(R)$ for the (nuclear) 
variables $R$.  Hunter, proposed accordingly that the molecular wavefunction 
$\Psi_{\rm mol}(r,R)$, i.e., the joint probability density amplitude of electronic 
and nuclear degrees of freedom could also be decomposed exactly as a product of a 
conditional probability amplitude $\Phi_R(r)$ of electronic variables, given fixed 
values for the nuclear variables times a marginal probability amplitude $X(R)$ for 
the nuclear degrees of freedom \cite{hunter75}:
 \begin{equation} \label{one}
\Psi_{\rm mol}(r,R)  = \Phi_R(r) \, X(R).
\end{equation}
The BO electronic and nuclear wavefunctions $\Phi^{\rm BO}_R(r)$, $X^{\rm BO}(R)$ 
provide excellent approximations for $\Phi_R(r)$ and $X(R)$ but Hunter suggested 
that for each $\Psi_{\rm mol}(r,R)$ one may find a conditional and a marginal 
probability density amplitudes, such that the factorization is exact. Czub and 
Wolniewicz discovered that for diatomic molecules, $X(R)$ is a nodeless wavefunction 
and hence the nuclear non-adiabatic potential must necessarily demonstrate spikes at 
the points were the adiabatic wavefunction has nodes \cite{czub}.  Hunter observed 
that this result is more general \cite{hunter80,hunter81}.  The proof is 
straightforward: Take a molecular wavefunction that is to a good approximation given 
by a single BO product, $\Psi_{\rm mol}(r,R) = \Phi^{\rm BO}_{1, R}(r) \, X^{\rm BO}(R)$, 
with the electronic part in the ground state and the nuclear wavefunction in an excited 
state with a node at $R_0$.  Since the electronic adiabatic states form a complete set, 
the molecular wavefunction can be expanded exactly in the series.
\begin{equation} \label{bhseries}
\Psi_{\rm mol}(r,R) = \sum_n \Phi^{\rm BO}_{n, R}(r) \, X_n(R)  .
\end{equation} 
Consequently, the diagonal of the nuclear $N$-body density matrix is equal to
\begin{equation}
\langle \Psi_{\rm mol}(R) | \Psi_{\rm mol}(R) \rangle =  \sum_n | X_n(R) |^2   .
\end{equation}
Here and in the following, integration over all electronic degrees of freedom is 
denoted by the bra-ket symbols, while integration over the nuclear degrees of 
freedom is written as $\int d^{3N}R$.\\ In order for 
$\langle \Psi_{\rm mol}(R) | \Psi_{\rm mol}(R) \rangle$ to have a node at $R_0$, 
all vibrational states $X_n(R)$ must have a node at the same $R_0$, which cannot 
happen. As an interesting aside, we note that Hunter's argument fails for nodes 
dictated by the symmetry.  For example for identical fermionic nuclei, the molecular 
wavefunction must be antisymmetric with respect to exchange of two nuclei, say at 
positions ${\bf R}_1$, ${\bf R}_2$ and then the diagonal $N$-body nuclear density 
matrix and consequently the marginal probability amplitude must have nodes at 
$({\bf R}_1, {\bf R}_2) = ({\bf R}, {\bf R})$.

So far, the impact of the work on the exact factorization of the molecular 
wavefunction has been limited and these ideas have not led to new practical ways to 
improve upon the adiabatic approximation.  Perhaps in the early papers, attention 
focused too heavily on the nodeless character of the marginal probability density 
amplitude, which admittedly is a non-adiabatic coupling but probably not a very 
important one, since when the BO approximation is accurate, the correction does not 
have a measurably significant effect and when the BO approximation breaks down, it 
is not due to the nodeless property of the wavefunction.  More importantly and rather 
curiously, the non-adiabatic equation determining the electronic wavefunction 
$\Phi_R (r)$ was never given. Knowledge of that equation would inform us of the 
non-adiabatic terms that correct the BO electronic wavefunction and hopefully accurate 
approximate treatments of these terms would emerge. This is the goal of our paper.  

There is a representability issue which perhaps was not addressed adequately, namely, 
it is not clear  whether the marginal probability distribution always corresponds to 
a meaningfull nuclear wavefunction.  For example the decomposition that was proposed 
to demonstrate the factorization (\ref{one}) with 
$\Phi_R(r)  =  { \Psi_{\rm mol}(r,R) / \sqrt{ \langle  \Psi_{\rm mol}(R)|\Psi_{\rm mol}(R) \rangle } }
$ and  $X(R)  =  \sqrt{ \langle  \Psi_{\rm mol}(R)|\Psi_{\rm mol}(R)\rangle } $

implies that the nuclear wavefunction is always bosonic even when the nuclei are 
fermionic.  This problem can be overcome by introducing an $R$-dependent phase and 
writing
\begin{eqnarray}
\Phi_R(r) & = & { |\langle g|\Psi_{\rm mol}(R)\rangle| \over \langle g|\Psi_{\rm mol}(R)\rangle }
{ \Psi_{\rm mol}(r,R) \over { \sqrt{\langle  \Psi_{\rm mol}(R)|\Psi_{\rm mol}(R) \rangle} } }
\\
X(R) & = & { \langle g|\Psi_{\rm mol}(R)\rangle \over | \langle g|\Psi_{\rm mol}(R)\rangle | } \, 
\sqrt{ \langle  \Psi_{\rm mol}(R)|\Psi_{\rm mol}(R)\rangle }   ,
\end{eqnarray}
where $g(r)$ is a suitably chosen electronic wavefunction that does not depend on $R$. 
In this way, the electronic wavefunction $\Phi_R(r)$ (conditional probability amplitude) 
depends on $R$ and for identical nuclei is symmetric with respect to nuclear exchange. 
The nuclear wavefunction $X(R)$ (marginal probability amplitude) has the same symmetry 
properties as $\Psi_{\rm mol}(r,R)$ with respect to nuclear exchange. 

Adopting the small potential mentioned earlier to confine the molecule in a very large 
volume, the molecular, electronic and nuclear wavefunctions are all normalized to unity 
for convenience:
\begin{equation} \label{molnorm}
\int d^{3N}R \langle  \Psi_{\rm mol}(R)|\Psi_{\rm mol}(R) \rangle = 1,
\end{equation}
\begin{equation} \label{elnorm}
\langle  \Phi_R | \Phi_R \rangle = 1,
\end{equation}
\begin{equation} \label{nunorm}
\int d^{3N}R |X(R)|^2 = 1.
\end{equation}

The electronic and nuclear wavefunctions have the freedom of an $R-$dependent phase 
(which, for identical nuclei must be symmetric with respect to exchange)
\begin{equation} \label{gauge}
\Phi_R (r)  \to e^{i \theta(R)} \, \Phi_R (r) ~,~~~ X(R) \to e^{-i \theta(R)} X(R)
\end{equation}
since this transformation leaves $\Psi_{\rm mol}(r,R) $ invariant. Apart from this 
freedom of phase the components $X$ and $\Phi_R$ are unique.

\section{Non-adiabatic electronic and nuclear equations}

As the parametric product ansatz (\ref{one}) does not constitute an approximation, 
it should be possible to improve significantly upon the BO level of accuracy, still 
using a product wavefunction (\ref{one}), if we optimize the electronic wavefunction 
variationally, under the normalisation constraint (\ref{elnorm}), rather than choose 
it a priori to be an eigenstate of the BO electronic Hamiltonian.  Naturally, we 
expect that in the non-adiabatic scheme, the electronic wavefunction will be coupled 
to the nuclear state.

The molecular Hamiltonian has the form:
\begin{equation} \label{h}
H_{\rm mol} = H^{\rm BO}_R -  \sum_{\alpha=1}^N {\hbar^2  \nabla_{\alpha}^2  \over 2 M_\alpha }
\end{equation}
where $\nabla_{\alpha}^2$ is the Laplacian with respect to ${\bf R}_{\alpha}$ and
$H^{\rm BO}_R$ is the electronic Hamiltonian:
\begin{equation} \label{hel}
H^{\rm BO}_R = - {\hbar^2  \over 2 m } \sum_{j=1}^{N_e} \nabla_{{\bf r}_j}^2
+ {1 \over 2} {\sum_{i, j =1 \atop i \neq j}^{N_e}  {1 \over r_{ij} }
- \sum_{j=1}^{N_e} \sum_{\alpha=1}^{N} { Z_\alpha \over |{\bf r}_j - {\bf R}_{\alpha}| }
+ {1 \over 2} {\sum_{\alpha, \beta=1 \atop \alpha \neq \beta}^{N}} 
{ Z_\alpha Z_\beta \over R_{\alpha \beta} } } .
\end{equation}
The nuclear Coulomb energy (last term) is independent of $r$ and represents a constant 
shift of the electronic energy.

In order to obtain the equations determining the electronic and  the nuclear 
wavefunctions, $\Phi_R(r)$ and $X(R)$, we have to vary the expectation value of 
the Hamiltonian $H_{\rm mol}$ with respect to $\Phi_R(r)$ and $X(R)$, under the 
normalisation constraints (\ref{elnorm}, \ref{nunorm}). The latter are incorporated 
through the use of Lagrangian multipliers $\Lambda '(R)$, $E$. Note we have infinitely 
many constraints for the normalisation of the electronic wavefunction, one for each 
$R$, and we need as many Lagrangian multipliers $\Lambda '(R)$:
\begin{equation} \label{lagrange 1}
- \int d^{3N} \! R \ \Lambda '(R) \, \left( \langle \Phi_R | \Phi_R \rangle - 1 \right)   ,
\end{equation}
\begin{equation} \label{lagrange 2}
- E \, \int d^{3N} R \, \big( |X(R)|^2 -1 \big)  .
\end{equation}
A few words about notation: If the gradient operator $\nabla_{\alpha}$, or the 
Laplacian $\nabla_{\alpha}^2$ lies within parentheses or between bra-kets, as in 
$\langle \Phi_R |\nabla_{\alpha} \Phi_R \rangle $, it is implied that it acts only 
inside the bra-ket or parentheses.  Note that when we use {\it square brackets}, as 
in $[ - i \hbar \nabla_\alpha \pm {\bf A}_\alpha ]$, the action of $\nabla_\alpha$ 
is not confined within the brackets. Direct optimization of the electronic and nuclear 
wavefunctions under the constraints (\ref{lagrange 1}, \ref{lagrange 2}) yields the 
non-adiabatic equations:
\begin{equation} \label{electronic}
\left[ H_R^{\rm BO}
- \sum_{\alpha=1}^N {\hbar^2  \nabla_{\alpha}^2 \over 2 M_\alpha }
+ \sum_{\alpha} { X^{\ast} (R) \left( - i \hbar \nabla_{\alpha}  X(R) \right) 
\over M_{\alpha} |X(R)|^2 } \cdot ( - i \hbar ) \nabla_{\alpha} \right]
\Phi_{R}(r) = \Lambda(R) 
\, \Phi_{R}(r)
\end{equation}
\begin{equation} \label{nucl2}
\left[  \sum_{\alpha=1}^N  {\big[ -i \hbar \nabla_{\alpha} +  
{\bf A}_{\alpha}(R) \big]^2 \over 2 M_\alpha } + 
\langle \Phi_R| H_R^{\rm BO} |\Phi_R\rangle + \sum_{\alpha=1}^N 
{ \hbar^2 \langle \nabla_{\alpha} \Phi_{R} | \nabla_{\alpha} 
\Phi_{R} \rangle  \over 2 M_\alpha} - 
\sum_{\alpha=1}^N {A_{\alpha} ( R)^2  
\over 2 M_\alpha } \right] \,  X(R) \nonumber \\
= E \, X(R)  .
\end{equation}
where $\Lambda(R) = \Lambda ' (R) / |X(R)|^2$ and $ {\bf A}_{\alpha} (R)$ is the 
vector-potential-like real quantity

\begin{equation} \label{quant}
{\bf A}_{\alpha} ( R) =  \langle \Phi_R| -i \hbar \nabla_{\alpha} \Phi_R \rangle   .
\end{equation}

To make further progress, we define an electronic energy as
\begin{equation} \label{X}
{\mathcal E}(R) =
\bigg\langle \Phi_R \bigg| H^{\rm BO}_R
+ \sum_{\alpha=1}^N  {[- i \hbar \nabla_{\alpha} - {\bf A}_{\alpha}(R)]^2
\over 2 M_{\alpha} } \bigg| \, \Phi_R \bigg\rangle
\end{equation}

which, by Eq. (\ref{quant}), can be written as
\begin{equation} \label{Y}
{\mathcal E}(R) =
\langle \Phi_R |H^{\rm BO}_R |\Phi_R  \rangle
+ \sum_{\alpha=1}^N  { \hbar^2  \langle \nabla_{\alpha} \Phi_R | \nabla_{\alpha} 
\Phi_R  \rangle \over 2 M_{\alpha} }
- \sum_{\alpha=1}^N  { A_{\alpha} (R)^2 \over  2 M_{\alpha} }   .
\end{equation}
After some straight-forward algebra, Eqs. (\ref{electronic}) and (\ref{nucl2}) can be 
expressed in terms of this electronic energy as
\begin{eqnarray} \label{electronic energy 1}
\lefteqn{
\left \{ H^{\rm BO}_R + \sum_{\alpha=1}^N { [- i \hbar \nabla_{\alpha} - 
{\bf A}_{\alpha}(R)]^2 \over 2 M_{\alpha} } \right.}
   \nonumber\\
&&\left. + \sum_{\alpha=1}^N {1 \over M_{\alpha} } \left [ { X^{\ast} (R) 
\left(- i \hbar \nabla_{\alpha} X(R) \right) \over  |X(R)|^2 }
+ {\bf A}_{\alpha}(R) \right] \cdot \Bigg [ - i \hbar \nabla_{\alpha} - 
{\bf A}_{\alpha}(R) \Bigg] \right  \}
\Phi_R (r)={\mathcal E}(R) \Phi_R (r)
\end{eqnarray}

\begin{equation} \label{electronic energy 2}
\left [\sum_{\alpha=1}^N { [- i \hbar \nabla_{\alpha} + {\bf A}_{\alpha}(R)]^2
\over 2 M_{\alpha}  }
+ {\mathcal E}(R) \right] X (R) = E X (R)   .
\end{equation}

While, from the numerical point of view, this form of the equations is clearly not 
advantageous, the new form reveals some other interesting aspects: First of all, Eqs 
(\ref{electronic energy 1}) and (\ref{electronic energy 2}) are manifestly form-invariant 
under the gauge transformation (\ref{gauge}). In particular, the electronic energy 
(\ref{X}) which appears as ``eigenvalue'' in (\ref{electronic energy 1}), is gauge 
invariant.  The third term on the left-hand side of Eq. (\ref{electronic energy 1}), 
while essential for the coupling between $X$ and $\Phi_R$, does not contribute to the 
energy ${\mathcal E}(R)$, i.e., when Eq. (\ref{electronic energy 2}) is multiplied from 
the left with $\Phi_R(r)^\ast$ and integrated over $r$, this term drops out.  One may 
confirm directly that the solution of the non-adiabatic equations (\ref{electronic}, 
\ref{nucl2}) is exact by showing that if $\Phi_{R}(r) $ and $X(R)$ satisfy 
(\ref{electronic}, \ref{nucl2}) then their product will be an eigenstate of 
$H_{\rm mol}$:
\begin{equation}
H_{\rm mol} \Phi_{R} (r) \, X(R) = E \, \Phi_{R} (r) \, X(R)  .
\end{equation}

If the phase of the electronic wavefunction $\Phi_{R}(r)$ depends on $r$ only, and if 
the dependence on $R$ is continuous and differentiable, then the vector potential 
${\bf A}_{\alpha}(R)$ vanishes and (\ref{nucl2}) simplifies to 
\begin{equation} \label{nucl3}
\left[ -  \sum_{\alpha=1}^N  {\hbar^2 \nabla_{\alpha}^2 \over 2 M_\alpha }
+ \langle \Phi_R| H_R^{\rm BO} |\Phi_R  \rangle
+ \sum_{\alpha=1}^N { \hbar^2 \langle \nabla_{\alpha} \Phi_{R} | \nabla_{\alpha} \Phi_{R} 
\rangle  \over 2 M_\alpha}
  \right] \,  X(R) = E \, X(R)   .
\end{equation}

The algebraic structure of the non-adiabatic nuclear equations (\ref{nucl2}, 
\ref{electronic energy 2}, or \ref{nucl3}), coincides with the nuclear equation one 
encounters in the adiabatic approximation.  However, in the latter, the electronic 
wavefunction is the BO one, satisfying
\begin{equation} \label{electronicBO}
H_R^{\rm BO} \, \Phi^{\rm BO}_R(r) = {\mathcal E}^{\rm BO}(R) \, \Phi^{\rm BO}_R(r),
\end{equation} 
with ${\mathcal E}^{\rm BO}(R) = \langle \Phi^{\rm BO}_R| H_R^{\rm BO} |\Phi^{\rm BO}_R 
\rangle $, while in the non-adiabatic nuclear equations (\ref{nucl2}, 
\ref{electronic energy 2}, or \ref{nucl3}), the electronic wavefunction $\Phi_R(r)$ is 
the non-adiabatic one satisfying (\ref{electronic}) or (\ref{electronic energy 1}).

The variational derivation of Eq. (\ref{nucl2}) guarantees that even in the approximate 
adiabatic scheme, the lowest energy eigenvalue, denoted by $E^{\rm BO}$, is rigorously 
an upper bound to the exact ground state energy of the system:
\begin{equation} \label{elebo}
E < E^{\rm BO}
\end{equation}
since $E$ is the minimum of the molecular Hamiltonian in a Hilbert space which is wider 
than the Hilbert space of the adiabatic states. The inequality holds even when the 
electronic non-adiabatic equation is derived in a constrained way, as in the following 
section on the Optimized Effective Potential Method treatment of the non-adiabatic terms.  
It is interesting to note that
\begin{equation} \label{erboler}
\langle \Phi_R |H_R^{\rm BO}| \Phi_R \rangle > {\mathcal E}^{\rm BO}(R)
\end{equation}
since ${\mathcal E}^{\rm BO}(R)$ is, for each $R$, the minimum of the expectation value 
of $H_R^{\rm BO}$.

A necessary consequence of (\ref{elebo}) and (\ref{erboler}) is that
\begin{equation}
\sum_{\alpha=1}^N { \hbar^2 \over 2 M_\alpha}  \, \int d^{3N} R \,  |X(R)|^2 \, 
\left(  \langle \nabla_{\alpha} \Phi_{R} | \nabla_{\alpha} \Phi_{R} \rangle  -
\langle \nabla_{\alpha} \Phi^{\rm BO}_{R} | \nabla_{\alpha} \Phi^{\rm BO}_{R} \rangle  
\right)  < 0
\end{equation}
where $X(R)$ is the lowest non-adiabatic nuclear state.  Observe that all terms in the 
inequality are nonnegative. The inequality indicates that ``on average'' for the 
non-adiabatic electronic state that corresponds to the lowest nuclear state $\langle 
\nabla_{\alpha} \Phi_{R} | \nabla_{\alpha} \Phi_{R} \rangle  < \langle \nabla_{\alpha} 
\Phi^{\rm BO}_{R} | \nabla_{\alpha} \Phi^{\rm BO}_{R} \rangle  $. So, $\Phi_{R}$ can be 
expected to change less rapidly than $\Phi^{\rm BO}_{R}$ as a function of $R$.  It is 
worthwhile to remember that when a nucleus samples a region of a degeneracy or an avoided 
crossing of electronic levels, the concept of the adiabatic surface looses its meaning and 
an indication is that the potential energy depends strongly on the nuclear position.  Small 
changes in $R$ result in large changes in the wavefunction and potential.  A lot of effort 
has been invested in constructing a different basis in the space of (near) degenerate 
electronic wavefunctions with energies that would behave smoothly close to the (avoided) 
level crossing.  A convenient choice is the diabatic basis \cite{smith}, defined as the 
basis that diagonalises the off-diagonal matrix elements 
$\langle \Phi_{R, m} | \nabla_\alpha \Phi_{R, n} \rangle$ \cite{smith,tru}. It is known 
however, that rigorously, such a basis does not exist in general \cite{tru,tru2}.  The 
non-adiabatic electronic wavefunctions and energies may offer a physical solution to this 
problem.  There are cases when the the vector-like-potentials ${\bf A}_{\alpha}$ cannot 
all be made to vanish (for example when adiabatic electronic levels become degenerate at 
a point $R$),  leading to a pseudo-magnetic field in (\ref{nucl2}) and a geometric phase 
\cite{mead}.  An interesting question is whether the geometric phase is a consequence of 
the approximate nature of the adiabatic approximation and consequently whether it would 
disappear in the exact non-adiabatic scheme.  We believe the geometric phase remains in 
the non-adiabatic scheme, since the non-adiabatic corrections depend on the nuclear mass 
and vanish for $M \rightarrow \infty $, whereas the geometric phase is independent of the 
size of the nuclear mass.  The geometric phase is then not an artefact of the adiabatic 
approximation but a result of the parametric decomposition (\ref{one}) of the molecular 
wavefunction which, by itself, is not an approximation.

\section{Approximations for the electronic non-adiabatic states}

The exact non-adiabatic electronic equation is naturally very hard to solve and one has 
to resort to approximation techniques.  A perturbative treatment of the non-adiabatic 
terms in the electronic equation is straightforward to apply if we keep the nuclear 
wavefunction $X(R)$ fixed.  However, when the non-adiabatic terms are large (which is the 
case of interest here), the expansion may converge very slowly, or not converge at all. 
Another method to deal with the non-adiabatic electronic equation, that seems particularly 
promising, is to seek for the local $R$-dependent potential $V_R({\bf r})$ that best 
approximates the effect of the non-adiabatic terms in (\ref{electronic}), in a manner 
analogous to the Optimised Effective Potential (OEP) method \cite{sh,ts,oep}.  

Remember that the non-adiabatic equation was derived under the constraint of 
normalisation of the electronic wavefunction (\ref{elnorm}). This was the most general 
way of requiring that the electronic wavefunction depends parametrically on the nuclear 
positions.  A more restrictive but perhaps more physical way is to impose that the 
electronic wavefunction must be an eigenstate of a Hamiltonian which is the sum of 
$H_R^{\rm BO}$ and an electronic potential $V_R$ that depends on the nuclear positions.  
The variationally best potential is determined by making the expectation value of the 
molecular Hamiltonian (\ref{h}) stationary.  The optimal $V_R$ will in general be an 
electronic $N_e$-body potential.  A further approximation, to make the problem tractable, 
is to restrict to (local) one-body potentials.  Hence we optimize the electronic 
wavefunction $\Phi_R(r)$ under the condition that it satisfies the equation
\begin{equation} \label{oepconstr}
\left[ 
H_R^{\rm BO} + \sum_{j=1}^{N_e} V_R({\bf r}_j)
\right] \Phi_R(r)
=
E(R) \, \Phi_R(r)  .
\end{equation}
In this way, the expectation value of the molecular Hamiltonian (\ref{h}) becomes a 
functional of the effective non-adiabatic potential, $\langle H_{\rm mol} \rangle = E[V_R]$.

Varying the effective potential, we obtain the functional derivative of $E[V_R]$ with 
respect to $V_R$:
\begin{eqnarray}
{ 1 \over |X(R)|^{2}} \ {\delta E[V_R] \over \delta V_R({\bf r})} & = &
\langle \Phi_R | {\hat \rho}({\bf r})  {\hat G}_R  {\hat V}_R  | \Phi_R \rangle 
+ \sum_{\alpha=1}^N {\hbar^2 \over 2 M_{\alpha} } \langle  \Phi_R | {\hat \rho}({\bf r}) 
{\hat G}_R
  | \nabla_{\alpha}^2 \Phi_R \rangle + {\rm c.c.}
\nonumber \\
&&
+ \sum_{\alpha=1}^N {\hbar^2 \over 2 M_{\alpha}} \, {  \big( \nabla_{\alpha} X(R) 
\big) \over X(R) } 
\cdot \langle \Phi_R | {\hat \rho}({\bf r})   {\hat G}_R | \nabla_{\alpha} 
\Phi_R \rangle + {\rm c.c.}
\label{2q}
\end{eqnarray}
where,
\begin{equation} \label{q2b}
{\hat G}_R = {\sum_n}' {| \Phi_{R, n} \rangle \langle \Phi_{R, n} |
\over {\mathcal E}_n(R) - {\mathcal E}(R) }  .
\end{equation}
Summation is over all eigenfunctions of, $H_R^{\rm BO} + V_R$, except $\Phi_R$. 
For compactness, we use in (\ref{2q},\ref{q2b}) second-quantization notation,  
${\hat \rho}({\bf r})$ is the electron charge density operator and 
${\hat V}_R = \int d^3{\bf x} V_R({\bf x}) {\hat \rho}({\bf x})$. Obviously, 
${\hat G}_R | \Phi_R \rangle = 0$.

The optimized effective potential we seek corresponds to a stationary point of 
$E[V_R]$, i.e. the functional derivative vanishes
\begin{equation} \label{oep}
{\delta E[V_R] \over \delta V_R({\bf r})} = 0.
\end{equation}

Solving the OEP equation (\ref{oep}) to obtain the potential can be very expensive.
The approximation proposed by Krieger, Li and Iafrate \cite{kli,kli2} (KLI) can be 
used to simplify the equation considerably.   

When the electronic state is a Slater determinant of Kohn-Sham spin-orbitals
$\varphi_{R,\sigma}^j$, one obtains the spin-dependent KLI effective potential:
\begin{eqnarray} \label{KLI}
\lefteqn{ 2 V_{R,\sigma}({\bf r}) \rho_\sigma ({\bf r})
=  2 \int d{\bf r}' V_{R,\sigma}({\bf r}')
\sum_j |\varphi_{R,\sigma}^j({\bf r})|^2 |\varphi_{R,\sigma}^j({\bf r}')|^2 }
 \nonumber \\
&& - \sum_\alpha {\hbar^2 \over 2 M_\alpha}  \sum_j
{\varphi_{R,\sigma}^j}^*({\bf r}) \nabla_{\alpha}^2 \varphi_{R,\sigma}^j({\bf r})
+ \sum_\alpha {\hbar^2 \over 2 M_\alpha}  \sum_j
|\varphi_{R,\sigma}^j({\bf r})|^2  \langle \varphi_{R,\sigma}^j| \nabla_{\alpha}^2 
\varphi_{R,\sigma}^j \rangle
+ {\mathrm c.c.} \nonumber \\
&& - \sum_\alpha {\hbar^2  \over M_\alpha} { \big( \nabla_\alpha X(R) \big) \over X(R) }
\cdot \sum_j \big[
{\varphi_{R,\sigma}^j}^*({\bf r}) \nabla_{\alpha} \varphi_{R,\sigma}^j({\bf r})
- |\varphi_{R,\sigma}^j({\bf r})|^2  \langle\varphi_{R,\sigma}^j| \nabla_{\alpha} 
\varphi_{R,\sigma}^j\rangle \big]
+ {\mathrm c.c.}
\end{eqnarray}
This equation can be solved within the ordinary Kohn-Sham self-consistency loop:
The potential $V_{R,\sigma}({\bf r})$ to be used in the next iteration is obtained by 
plugging the Kohn-Sham orbitals and the $V_{R,\sigma}({\bf r})$ of the previous 
iteration in the right-hand side of Eq. (\ref{KLI}).

We have solved the non-adiabatic OEP equation for the simplest system that contains 
non-adiabatic couplings, namely the hydrogen ion $H_2^+$. Using the BO scheme one 
obtains an excellent approximation for the ground state wavefunction, 
$\Psi({\bf r},{\bf R}) = \Phi_{\bf R}^{\rm BO}({\bf r}) \, X^{\rm BO}({\bf R})$.
To test our method, we used the inverse BO separation, i.e., we wrote,
$\Psi({\bf r},{\bf R}) = \phi_{\bf r}^{\rm iBO}({\bf R}) \, \chi^{\rm iBO}({\bf r})$, 
where $\phi_{\bf r}^{\rm iBO}({\bf R})$ is the conditional probability density amplitude 
for the relative nuclear coordinate, when the electron is frozen at $\bf r$, and 
$\chi^{\rm iBO}({\bf r})$ the marginal probability density amplitude for the electron, 
moving in the potential surface of the nuclei. The inverse BO separation ignores terms 
of the order $M/m$ and is hence very poor. Including the non-adiabatic correction terms 
(in this case the non-adiabatic terms can be transformed to have the form of a potential 
and hence the OEP method is exact), we found that the solution converged {\it numerically} 
to the correct wavefunction one obtains doing the BO seperation in the orthodox way.

\section{Summary}
We have revisited the adiabatic/BO separation of the electronic and nuclear degrees of 
freedom in molecules, aiming to improve accuracy beyond the adiabatic level, but keeping 
at the same time the language and concepts of the adiabatic approximation which form the 
basis of almost all studies of molecular systems. We have derived a non-adiabatic 
correction for the electronic equation, which couples in a self-consistent way the 
electronic and nuclear wavefunctions. This correction has the form of an effective 
potential. The implementation of this correction to electronic structure codes that 
already use the OEP method to account for electronic exchange and correlation seems 
particularly promising. 

\section{Acknowledgments}
This work was supported by the Deutsche Forschungsgemeinschaft within the 
Sonderforschungsbereich SFB450, by the EXC!TiNG Research and Training Network of the 
European Union, and by the NANOQUANTA Network of Excellence.

\end{document}